%% file: main.tex
\documentclass[runningheads]{llncs}
\usepackage{graphicx}
\usepackage{tikz}
\usetikzlibrary{arrows.meta}
\usepackage{booktabs}
\usepackage{array}
\newcommand{\secpagemark}[1]{\message{^^JSECPAGEMARK #1 @ \thepage^^J}}

\begin{document}

\title{SoK: Failure Modes in Common Criteria Product Evaluation---A Taxonomy and Design-for-Evaluability Guidance}
\titlerunning{SoK: Failure Modes in Common Criteria Product Evaluation}
\author{Punit Suketu Patel\orcidID{0009-0007-8575-0765}}
\authorrunning{P. S. Patel}
\institute{Independent Researcher, Dallas, TX, USA\\
\email{ppatel34@terpmail.umd.edu}}
\maketitle

\input{body}

\secpagemark{References}
\bibliographystyle{splncs04}
\bibliography{references}

\input{appendixA}

\end{document}

%% file: body.tex
\begin{abstract}
The Common Criteria (CC; ISO/IEC 15408) is the principal international framework for evaluating the security of IT products, and its certificates gate procurement across government, defense, and regulated industry. Yet evaluations routinely stall, fail, or yield certificates whose assurances do not survive real-world deployment. Prior work has approached this from two directions: economic critiques of misaligned incentives and ``security theatre''; and recent data-driven studies that quantify vulnerabilities in already-certified products. Comparatively little systematizes the problem from the evaluator's operational vantage --- recurring, cross-vendor failures against the Common Criteria work units themselves. This paper presents a systematization of knowledge (SoK) of failure modes in Common Criteria product evaluation. Drawing on the standard, the Common Evaluation Methodology (ISO/IEC 18045), public NIAP Protection Profiles, and published Security Targets and Certification Reports, we organize recurring failures into a lifecycle taxonomy spanning Security Target scoping, Protection Profile conformance, assurance evidence, cryptographic requirements, functional testing, vulnerability analysis, operational guidance, and post-certification configuration drift. We analyze their root causes, derive a design-for-evaluability framework that product teams can apply before evaluation begins, and identify open problems for cloud, continuous-delivery, and AI-enabled systems.

\keywords{Common Criteria \and security certification \and security evaluation \and failure taxonomy \and design for evaluability}
\end{abstract}

\section{Introduction}
\secpagemark{Introduction}

When a government agency buys a firewall, a router, or an operating system, it cannot practically audit the product's internals itself. It relies on certification instead. The Common Criteria (CC; ISO/IEC 15408) is the international standard behind that reliance: an accredited laboratory evaluates the product against defined security requirements, a national scheme reviews the laboratory's work, and the resulting certificate is recognized by the more than thirty member countries of the Common Criteria Recognition Arrangement \cite{ccra2014}. In the United States, certification against Protection Profiles published by the National Information Assurance Partnership (NIAP) gates procurement for national security systems \cite{cnssp11}. A certificate is not a formality. It decides what can be sold, and to whom.

The process behind that certificate is neither quick nor cheap. Evaluations routinely take months to years and cost hundreds of thousands of dollars \cite{janovsky2024seccerts}. It is fair to ask what the money buys, because the record is uncomfortable. Critical vulnerabilities such as ROCA, TPM-Fail, and Minerva have been discovered in products certified at high assurance levels \cite{nemec2017,moghimi2020,jancar2020,janovsky2024seccerts}. A certified PIN-entry device was defeated by cutting a small hole in its case and hooking a paperclip onto an internal communication line \cite{murdoch2012}. Evaluations also fail in a quieter way: they stall, overrun, and consume budgets on problems that better preparation would have prevented. Between ``conformant on paper'' and ``secure as deployed'' there is a gap wide enough that serious researchers have asked whether certification amounts to security theatre \cite{anderson2009}.

Prior work has examined this gap from two directions. Security economists explain it institutionally: the vendor chooses and pays the laboratory, and a market for lemons follows \cite{anderson2009,murdoch2012}. Measurement researchers mine the public corpus of certificates to map disclosed vulnerabilities onto certified products at scale \cite{janovsky2024seccerts}. A third strand surveys the standard itself and the barriers to adopting it \cite{sun2022}. We build on all three. But each treats the evaluation as a black box. Documents and money go in; a certificate comes out; the analysis happens on either side.

What happens inside the box is where evaluations actually succeed or fail. An evaluation is not an abstraction. It is a person in a laboratory unboxing a router, following the vendor's guide to reach the exact configuration that is supposed to be evaluated, deliberately building broken certificate chains, and watching a packet capture to see whether the device rejects them for the right reason. From that chair, failure has a texture that neither the economics nor the measurement literature captures: the setup guide that cannot reproduce the evaluated configuration; the claimed cipher suite that never appears in a handshake; the entropy assessment that demands more raw data than the device can physically store. These problems recur across vendors and product categories. They are systematic. They have not, so far, been systematized in a form that evaluators and product teams can act on.

That is what this paper does. We draw on the standard, public NIAP Protection Profiles, published Security Targets and Certification Reports, and the literature above, combined with the author's experience performing Common Criteria evaluations at a NIAP-accredited commercial testing laboratory. That experience runs deepest in cryptographic functional testing and in the evidence and guidance work that surrounds it. The firsthand experience contributes the vantage, the emphasis, and the fixes.

The paper makes four contributions.

\textbf{C1~--- Taxonomy.} A lifecycle taxonomy of nine classes of CC evaluation failure modes (E1--E9), organized by evaluation stage and Common Evaluation Methodology (CEM) work unit. Where prior economic critiques treat certification failure at the institutional level, we operationalize it at the level of the evidence and activities an evaluator actually assesses.
\textbf{C2~--- Root-cause analysis.} For each class, the underlying cause (scoping decisions, evidence gaps, product-vs-process mismatch), surfacing cross-cutting patterns such as the ``conformant-on-paper vs. secure-as-deployed'' divergence.
\textbf{C3~--- Design-for-evaluability framework.} Concrete, taxonomy-linked practices that product teams can adopt during design and development to prevent the identified failures and shorten evaluation --- turning a largely diagnostic literature constructive.
\textbf{C4~--- Open problems.} Challenges for CC as products move to cloud/SaaS, continuous delivery, and AI/ML components, where the fixed-configuration, point-in-time evaluation model strains.

The remainder of the paper presents the taxonomy (Section 4), its cross-cutting patterns (Section 5), the design-for-evaluability framework (Section 6), open problems (Section 7), and related work (Section 8).

\section{Background}
\secpagemark{Background}

The Common Criteria relies on a specific, often impenetrable vocabulary; this section defines the essential terms and clarifies what a certificate actually claims.

\subsection{The cast of documents}

The \textbf{Target of Evaluation (TOE)} is the part of the product inside the evaluation boundary, with everything outside it delegated to the operational environment --- a line Section 4 returns to. The \textbf{Security Target (ST)} is the vendor's claim document, and an evaluation is always against an ST: nothing more is examined, and nothing less. \textbf{Protection Profiles (PPs)} package a category's \textbf{Security Functional Requirements (SFRs)} and assurance requirements. Historically, evaluations were graded by \textbf{Evaluation Assurance Levels (EALs)}; the United States scheme has since retired EALs in favor of \textbf{exact conformance}, every product in a category evaluated against the same PP, no more and no less \cite{niapndcpp}. This paper covers both styles, with emphasis on the PP-based model. Finally, the \textbf{Common Evaluation Methodology (CEM; ISO/IEC 18045)} defines, for each assurance requirement, the concrete \textbf{work units} an evaluator must check and document. When this paper's taxonomy points at a work unit, it is pointing at the CEM. (The 2022 revision, CC:2022, reorganizes the documents and ratifies this model: exact conformance enters the standard itself, a new Part 4 formalizes evaluation activities, and the work units persist.)

\subsection{How an evaluation proceeds}

The vendor writes the ST and builds the evidence. The laboratory's work is where this paper lives: evaluators bring the product to its \textbf{evaluated configuration} by following the vendor's own guidance documents, run the prescribed tests, perform vulnerability analysis, and record evidence for every verdict. The output is an \textbf{Evaluation Technical Report (ETR)} for the scheme's validators; the ST and a \textbf{Certification Report} are published, and the ETR is not.

\subsection{What a certificate does and does not claim}

A certificate makes a narrow claim: this specific product version, in this specific evaluated configuration, was examined against the requirements in its ST, following the CEM, by an accredited laboratory, and a national scheme reviewed that work. Every word of that sentence is load-bearing.

It does not claim the product is free of vulnerabilities. It does not cover features outside the TOE boundary, or configurations other than the evaluated one, and it says nothing about the next software version; the certificate is a snapshot of a fixed version at a point in time. Nor is a certificate a general endorsement. A product can be certified while its most-used features were never examined, if those features sat outside the boundary the ST drew.

The evaluated configuration deserves particular attention, because so much later trouble traces back to it. The vendor's guidance documents are supposed to lead a competent administrator, step by step, to the exact configuration that was tested. If the documentation cannot do that, the assurance exists on paper but cannot be reproduced in the field. And if an operator runs the product outside that configuration, the certificate's claim simply does not apply. Sections 4 and 7 return to both problems.

Held against these boundaries, the failures catalogued in Section 4 stop looking like paradoxes and start looking predictable.

\section{Sources, scope, and confidentiality}
\secpagemark{Sources-scope-and-confidential}

This systematization draws on four kinds of sources: the normative record (the standard, the Common Evaluation Methodology, and the NIAP Protection Profiles with their per-requirement test activities); the public evaluation record (the Security Targets and Certification Reports published on the Common Criteria portal and the NIAP Product Compliant List); and the literature of Sections 1 and 8. The fourth is firsthand: the author performed Common Criteria evaluations of network and server products at a NIAP-accredited commercial testing laboratory, with work concentrated in cryptographic functional testing, evidence collection, and guidance verification.

The taxonomy was not assembled from a blank page. Its nine classes follow the path an evaluation actually walks, from scoping the Security Target to what happens after the certificate is issued, and each class is anchored to the assurance family or CEM work units where that failure surfaces. A failure mode earned a place in the taxonomy if it met three tests: it recurs across vendors and product categories rather than being one product's bad day; it maps to an identifiable work unit, so an evaluator can point to where it bites; and it is actionable, meaning a product team could have prevented it before evaluation began. Depth of treatment deliberately follows depth of evidence. The classes closest to the author's hands-on work receive the most detailed treatment; those further from it are systematized from the public record and literature, and the text says so where it matters.

Evaluation laboratories operate under strict client confidentiality, and this paper honors it: no client engagement, vendor identity, test verdict, or laboratory-internal artifact from the author's work appears here. Every product-specific failure is drawn from published sources; the firsthand experience appears only as generalized descriptions of evaluation practice that any evaluator at any accredited laboratory would recognize. The boundary is an ethical obligation but also a methodological one: every checkable claim in this paper can be checked.

Three limitations should be stated plainly. This is a qualitative systematization: it identifies recurring failure classes and explains their mechanics but makes no claims about relative frequency; the certificate-corpus line is the right instrument for rates \cite{janovsky2024seccerts}, and the two approaches are complementary. The firsthand vantage is one evaluator's, at one laboratory, under one scheme, in the 2019--2021 window; we mitigate it by anchoring every class to scheme-independent work units and grounding every named instance in the public record (Appendix~A), but readers should weigh it for what it is. And the standard is mid-transition to CC:2022; where the revision changes something this paper relies on, the text notes it.

\section{A taxonomy of evaluation failure modes}
\secpagemark{A-taxonomy-of-evaluation-failu}

\begin{figure}[t]
\centering
\input{fig1}
\caption{The nine failure classes (E1--E9) across the evaluation lifecycle.
Unshaded classes trace to decisions a product team controls and are addressed
by the practices of Section~6; shaded classes (E7, E9) trace to the evaluation
model itself and are taken up as open problems in Section~7.}
\label{fig:lifecycle}
\end{figure}
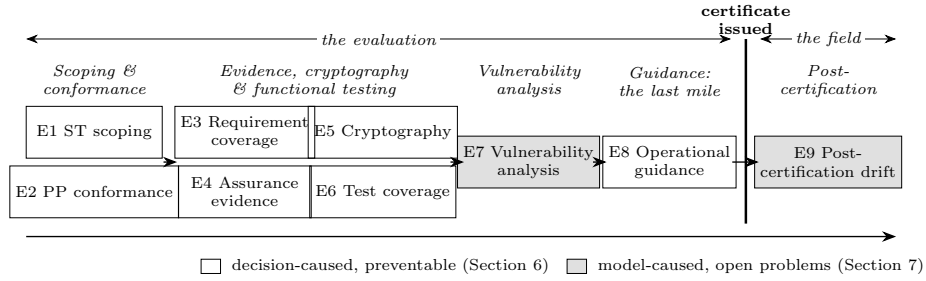

The classes below follow the life of an evaluation in the order the work happens. Figure~\ref{fig:lifecycle} places them on that lifecycle. Each class describes the failure, where it surfaces in the evaluation, what typically causes it, and what it costs. Prevention is collected separately in Section 6. Table~\ref{tab:modes} summarizes all nine.

\subsection{E1. Security Target scoping: mis-aimed before it starts}

Every downstream activity inherits from the Security Target: the boundary it draws decides what gets tested, what the guidance must cover, and what the certificate claims. E1 is the class of failures where that boundary is drawn wrong.

It takes three recurring shapes. A boundary drawn too broad commits the vendor to producing design evidence and test access for components it does not control (a bundled third-party library), and the evaluation stalls waiting for evidence nobody can produce. A boundary drawn too narrow produces a certificate that quietly excludes the parts customers actually rely on, a possibility Section 2.3 flagged and buyers rarely check. And the subtlest shape: security functions delegated to the operational environment. An ST can declare that authentication is handled by an enterprise directory, and the function becomes an assumption rather than an evaluated claim. Each delegation may be individually reasonable. Enough of them hollow the evaluation out.

The evaluator meets E1 in the ASE work units \cite{cem2022}, and the failure announces itself as churn: boundary redrawn, ST re-issued, test plans reworked. The root cause is usually organizational rather than technical. Scope gets set by product and marketing pressure (certify as much as possible) or by a consultant optimizing for the smallest evaluable claim \cite{anderson2009,arnold2006}, and in both cases before anyone asks the only question that matters: can we produce evidence for everything inside this line? Discovered late, a scoping mistake is among the most expensive to fix, because everything already built on the boundary moves with it.

The churn leaves public traces. One published Security Target's revision history records ``Expanded the TOE boundary to address certifier ORs'' \cite{st3837138}; another's logs the boundary moving in both directions during the evaluation, requirements removed in one revision and TOE components added in a later one \cite{bsi0531}.

\subsection{E2. PP conformance: claiming a bar the product cannot clear}

Under NIAP's exact-conformance model, an ST must match its Protection Profile requirement for requirement, with selections chosen only from what the PP permits \cite{niapndcpp}. E2 is the class of failures where the product cannot deliver what the claimed PP demands.

Sometimes a mandatory requirement is missing: the PP requires local audit storage, the appliance has none \cite{td0592}. Sometimes the vendor makes a selection the shipped build cannot honor: claiming a mutual-authentication option its TLS server cannot perform, or an algorithm the firmware never implemented \cite{td0009}. And sometimes the mismatch is architectural: the PP's assumptions describe a self-contained appliance while the product is an agent that depends on a host platform, and no amount of ST wordsmithing closes that gap.

E2 surfaces early in the conformance-claim checks (ASE\_CCL), but its fixes require product changes --- the slowest kind to implement. A missing feature cannot be argued into existence; the evaluation waits for a firmware release, the claim set is renegotiated, or the effort is abandoned. The root cause is almost always sequencing: the vendor chose the PP for market access before anyone ran a gap analysis against the exact PP version. Section 6 returns to the gap analysis as the cheapest insurance an evaluation can buy.

\subsection{E3. Requirement coverage: the wire disagrees with the paper}

E3 is the class where the claims are conformant, the boundary is sound, and the product still does not do what the ST says. The claim exists on paper; the behavior on the wire is different.

From the bench this is concrete. A Security Target lists twelve cipher suites; the evaluator enumerates the TLS handshake and finds nine. A version claim says older protocols are rejected; the device accepts one silently. Certificate validation is claimed in full; a deliberately broken chain --- an invalidated intermediate --- validates anyway. This is the ``conformant on paper, different in practice'' divergence at the level of individual requirements, and the standard test activities are designed to catch exactly this: offer one algorithm at a time, watch the handshake, compare what is negotiated against what is claimed \cite{niapndcppsd}.

How does an ST end up claiming behavior its product lacks? Rarely by intent. STs are copied forward from a prior product or version, and stale claims ride along. Claims are written from specification sheets rather than from the build that ships to the laboratory, and the two drift. And sometimes the claimed behavior exists but is unreachable in the evaluated configuration: the hardening guide that establishes the certified mode disables the very feature the ST claims.

The cost profile is what makes E3 painful: it surfaces mid-testing, after contracts, after scoping, after conformance review, the most expensive place to discover a false claim. Testing halts on the affected requirements, the ST is amended \cite{st3834314,stvid11253}, and amended STs re-open review. The prevention, developed in Section 6, is almost embarrassingly direct: run the laboratory's enumeration against your own build before the laboratory does.

\subsection{E4. Assurance evidence: the verdict is only as good as its record}

The output of an evaluation is not a pass or a fail. It is the record that justifies one: what was examined, on which build, with what result, assembled into the ETR. E4 is the class of failures where that record breaks down, and it has a vendor side and a laboratory side.

On the vendor side, the assurance requirements demand documents with specific jobs. A functional specification (ADV\_FSP) must describe the product's security interfaces precisely enough for the evaluator to check each one against the ST's claims. Configuration-management evidence (ALC\_CMC, ALC\_CMS) must identify the evaluated version uniquely, so that everyone knows the build on the bench is the build on the certificate. (Guidance documents earn their own class, E8.) The recurring failure is documents written for the wrong job \cite{arnold2006,ewa2006}. A functional specification assembled from datasheets describes what the product is for, not what its interfaces do. Version evidence names a number but cannot connect it to the firmware image delivered for testing. These failures do not announce themselves; they surface as iteration, each document bounced between laboratory and vendor until it can carry the checks that the CEM makes against it \cite{jtsecfaq}.

On the laboratory side the burden inverts: every verdict from the independent testing (ATE\_IND) needs evidence that would let a skeptical validator reconstruct what happened. In practice that means, per test case, a packet capture and screenshots tied to the exact test, build, and configuration. Multiply that across hundreds of test activities and several devices in flight at once, and evidence management becomes an engineering problem in its own right. The working answer is unglamorous: a directory tree that mirrors the report's structure, so that assembling the ETR is a walk through the tree rather than an archaeology dig. When that discipline slips, or when a mid-evaluation build change invalidates weeks of captures, evidence rework quietly becomes the schedule.

One evidence problem is structural rather than organizational. A successful test is easy to evidence: the handshake completed, and the capture shows it. A correctly failed test on an encrypted channel is not. Reject a wrong SSH password and the capture shows an encrypted exchange and a closed connection; the reason is invisible by design, and the terminal does not even echo the password that was typed. The record can establish that a rejection happened, but not that it happened for the right reason, so the strength of the evidence rests partly on the evaluator's procedure notes rather than on the artifacts themselves. Evidence of correct failure is structurally harder to produce than evidence of success. Section 5 develops this as a pattern.

E4's cost is rarely a failed evaluation. It is drag: document iterations, re-captures, validator comebacks when the record does not line up. The root cause on both sides is the same: treating evidence as a byproduct of the work rather than a product of it. Section 6's remedies are concrete: write the documents from the build rather than ahead of it, and decide before testing starts how each test's outcome will be observed and recorded.

\subsection{E5. Cryptography: exact demands on inherited machinery}

Cryptography holds a peculiar position in an evaluation. The FCS requirements are the most exact in any Protection Profile: named algorithms, specific modes and key sizes. There is no interpretive room; AES-GCM either is or is not what the device computes. Yet the cryptography is usually the part of the product its vendor built least. The algorithms come from a library, the randomness comes from silicon. E5 is the class of failures that grows in that gap: the ST's most precise claims attach to the components the vendor understands least.

The first shape is validation. Under NIAP policy, every claimed algorithm for which NIST runs a validation program must hold a CAVP certificate \cite{niappolicy5}. The recurring failures are mismatches of scope: a certificate that covers the library but not the mode or key size the ST selects, was earned on a different processor or operational environment than the TOE's, or predates the patch in the shipping binary. Each of these looks like a paperwork problem, and each is a hard stop: the laboratory cannot test an unvalidated algorithm into acceptability.

The second shape is entropy, where the standard's assumptions meet hardware reality hardest. Every key the device will ever generate inherits its strength from the random-bit generator, and the generator from a physical noise source. The PP therefore requires an entropy assessment \cite{sp80090b,niapndcppsd}: where the randomness physically comes from, health tests, and a min-entropy estimate from raw noise-source output. These requirements fail. The noise source often sits in licensed third-party silicon that the product vendor cannot document and the silicon supplier will not. Shipping firmware frequently has no supported way to expose raw noise output. And the analysis needs raw data at volumes that embarrass an embedded device: far more consecutive output than the device has storage to hold. From the bench, that last constraint turns into logistics: collection scripts that rotate, compress, and ship chunks of output off the device, so that the dataset accumulates on laboratory storage the device itself never had. The assessment gets done. But when gathering the evidence requires improvised plumbing, the requirement's assumptions have drifted far from the hardware it governs, a gap Section 7 takes up as an open problem.

The third shape is the end of a key's life. FCS\_CKM requires that keys be destroyed when no longer needed, and destruction must be demonstrated, not asserted. Modern storage resists the demonstration. On flash with wear-leveling, ``we zeroize the key'' can be true at the API and false on the medium; the cPP community has worked through public iterations on this exact requirement \cite{td0057,td0130}. The vendor is asked to prove a negative, that no copy survives, on hardware never designed to support the proof.

The root cause across all three shapes is the same: inheritance without ownership. Vendors claim what the library supports rather than what they have configured and verified. The product organization cannot describe a noise source it did not build. A single ST vouches for a chain that crosses two or three companies, and the evaluation is where that chain gets pulled taut. The cost profile follows: E5 failures tend to be stalls rather than quick verdicts, because their fixes wait on outsiders. Section 6's advice is about sequencing: inventory the cryptography before claiming it, and start the entropy work first.

\subsection{E6. Functional test coverage: tested where it is easy, assumed where it is hard}

On paper, the PP-based model has no coverage problem. Every claimed requirement comes with prescribed test activities, the evaluator runs them all, and coverage is nominally complete \cite{niapndcppsd}. E6 is the class of failures in the distance between that nominal completeness and what actually gets exercised: not false claims (that is E3), but true claims whose testing touches less of the product than the certificate implies. EAL-graded evaluations meet the same distance from the other side: there the vendor writes the test plan, and coverage is measured against the vendor's own functional specification.

The clearest shape is role asymmetry. Protocol requirements typically apply to a device in two roles, and the two are not equally testable. With the TOE as a TLS or SSH server, the initiating side is the evaluator's own machine, carrying every tool the test needs; driving the handshake, varying one parameter at a time, and capturing everything is routine. With the TOE as a client, the device must initiate the connection, from its own command line, with parameters the test controls, and whether that is even possible depends on what the product exposes. From the bench, exercising client behavior sometimes meant copying the test tooling onto the device and running it there, improvising the instrumentation the product never offered. The result is systematic: coverage quietly thins where the work is hardest. The certificate does not record the difference. A pass is a pass.

The second shape is behavior that cannot be induced. Some requirements are promises about failure: the noise source degrades and the health tests catch it; a self-test fails and the device refuses to start \cite{td0190}. Testing them means making the bad thing happen, and often nothing outside the device can make it happen, so the test leans on a vendor-supplied test build or debug hook: what gets exercised is a build modified to be testable, one step removed from the product that ships. The coverage exists formally. What it covers has drifted.

The third shape is that the tests form a list, not a matrix. The same one-requirement-at-a-time discipline that makes E3's false claims catchable also bounds what testing sees. Each cipher suite is offered alone; each policy is toggled alone. Interactions between requirements (session resumption meeting revocation checking) are prescribed nowhere: too specific for the per-requirement activities, and not the job of the survey-level vulnerability analysis that these evaluations run (E7). Combination behavior falls between the two.

None of this is evaluator laziness, and little of it is vendor deceit. Test effort follows testability, and testability was fixed by design decisions made years before anyone thought about evaluation. The PP's prescriptions are written for an entire product category, so they bend toward what every product can do, and the behaviors that are hard to reach stay hard to reach.

The cost lands in two places. During the evaluation it is time: rigging what the product should have provided. After certification it is invisible: assurance reads as uniform across the claimed requirements, but it is thickest where testing was easiest, and nothing on the certificate says which requirements got the thick end. Section 6's response is the most direct design-for-evaluability case in the paper: give the device a supported way to act as a client and a supported way to induce failure conditions, so that the hard half of the test surface costs what the easy half does.

\subsection{E7. Vulnerability analysis: what ``someone tried to break it'' actually means}

Of everything a certificate implies, nothing gets read more generously than the vulnerability analysis. A buyer told that certified products undergo vulnerability analysis and penetration testing will picture skilled attackers spending weeks trying to break the product. What the assurance family (AVA\_VAN) mandates in PP-based evaluations is narrower: a search of public sources for known vulnerabilities applicable to the TOE, followed by penetration testing of the resulting hypotheses, against an attacker of basic attack potential \cite{cem2022}. Every qualifier in that sentence removes something the buyer imagined, and E7 is the class of failures living in that removal. (It is also the class furthest from the author's hands-on work.)

The quietest qualifier is the adversary. Basic attack potential describes someone with limited time, standard equipment, and public knowledge; the products that seek certification attract exactly the attackers that description excludes. At the shallow end, the analysis never claimed to resist the adversary the product will actually face. At the deep end the news is harsher: ROCA was found in chips certified at high assurance levels, where the analysis did claim resistance to high attack potential, and sophisticated cryptanalysis walked past it anyway \cite{nemec2017,janovsky2024seccerts}. The analysis bounds what the certificate can promise, and novel discovery respects no bound.

The survey is also keyed to what is known and what is named. A vulnerability published after the survey date is invisible to it: the certificate is a snapshot, and E9 returns to what happens next. And public-source searching runs on names, so a device carrying an outdated TLS library inherits that library's vulnerability history, but a survey that does not know the library is there cannot look it up.

And unlike everything else in the evaluation, this activity resists prescription. Functional testing is specified step by step (E3, E6); vulnerability analysis can be specified only as a process: search, hypothesize, test. There is no work unit for finding an unknown flaw. Depth therefore rides on the evaluator's skill, curiosity, and remaining calendar, and where depth is discretionary and the vendor chooses the laboratory, discretion feels commercial gravity \cite{anderson2009,murdoch2012}. The PIN-entry device of Section 1 is the canonical instance: nobody had hypothesized anything so cheap \cite{murdoch2012}.

The root cause is a mismatch of kinds: discovery is open-ended, evaluation is bounded. Schemes bound the activity the only ways they can (by attack potential, by public sources, by time), and each of those lines is one that a real attacker does not observe. The survey still has value: fed an honest component inventory, it reliably catches inherited CVEs and cheap flaws, which is exactly the low bar certified products have publicly failed to clear. But the limits appear nowhere a buyer will look.

The cost of E7 is unusual in this taxonomy: it arrives too late and spreads too thin to change anyone's behavior. A shallow analysis rarely stalls the evaluation that contains it; it surfaces years later as a certified-yet-vulnerable headline, and the damage lands on parties who had no say in the analysis's depth: buyers who trusted the certificate, and the scheme whose credibility absorbs the security-theatre critique that each new instance revives \cite{anderson2009}. Section 6's fixes are cheap and vendor-side: disclose the component inventory, sweep it for CVEs before the laboratory does, and buy deeper adversarial testing separately when the threat model demands it.

\subsection{E8. Operational guidance: the certificate's last mile}

Guidance is the one assurance deliverable that leaves the laboratory. The guidance documents (AGD\_PRE, AGD\_OPE) ship with the product, and they carry the entire weight of Section 2.3's warning: the guidance is the only bridge from a device in a box to the evaluated configuration. E8 is the class of failures where the bridge does not carry.

The first shape is guidance that cannot get you there. Evaluation begins with bring-up: unbox the device and follow the vendor's own documents toward the evaluated configuration. This makes the evaluator the guidance's first field test, and the test does not always pass. Steps are missing or out of order; the menu in the manual was renamed two firmware versions ago. Deep enough into a wrong turn, the only way out is a factory reset. This happens to an evaluator holding the Security Target and the test plan, the documents that describe exactly what this device is supposed to become. If a reader that well equipped sometimes ends at a factory reset, the field administrator with a change window and a search engine cannot be expected to do better.

The second shape is a certified state that does not keep. Test activities have the evaluator configure a setting, reboot the device, and verify it is still enforced \cite{niapndcppsd}, because on real hardware the answer is not always yes. Guidance that omits the persistence step delivers an administrator to the evaluated configuration and lets the first reboot take it away. Nothing looks wrong; the device works; the certificate's precondition is simply no longer true.

The third shape is that the guidance that matters is often the guidance nobody reads. The evaluated configuration lives in a dedicated Common Criteria supplement, separate from the mainstream administration documents. The separation is understandable (certified mode is a restricted posture), and it is also how the last mile fails at scale: field administrators configure the product from the documents they know, and can run a certified product for years without learning that a CC supplement exists. Where the two document sets diverge, the certificate describes a configuration almost nobody runs. E9 begins here.

The root causes are about authorship. Guidance is commonly written away from the hardware: by technical writers or certification consultants \cite{arnold2006,rosenberry}, from specifications, before anyone has walked a fresh device through the procedure end to end. Documentation is not tested the way code is tested, and the guidance iterates mid-evaluation until following it produces the configuration it promises. Heterogeneity multiplies the cost: every vendor writes documentation in its own idiom, so fluency with one manufacturer's manuals is worth little on the next, a learning curve paid again by every evaluator and every customer.

E8's cost during the evaluation is time: bring-up, resets, guidance iterations. Its cost after certification is the gap Section 2.3 named: assurance that was real in the laboratory and unreachable in the field. Section 6's fixes are procedural and cheap: a fresh-device walkthrough of the draft guidance, persistence made explicit at every step that needs it, and a way for the device to answer the only question an administrator actually asks: am I in the evaluated configuration right now? A supported status command that attests the certified posture would do more than any amount of better prose.

\subsection{E9. Post-certification drift: the certificate stands still}

Every failure class so far lives inside the evaluation. E9 lives after it, in territory the evaluator never sees: it begins the day the certificate is issued, when the snapshot that Section 2.3 described starts to age. From that day, two clocks run against the certificate at once: the product's and the deployment's.

A certificate names an exact version, and the day the vendor ships a security patch, every operator of the certified product faces a choice between two goods: apply the patch and leave the certified version behind, or stay conformant and keep running code with a known vulnerability. Assurance-continuity processes exist to carry a certificate across minor updates \cite{ccraac2012,niapndcpp}, but maintenance moves at scheme speed while patches move at exploit speed, and certified fleets live in the gap: some patched out of conformance, some conformant and exposed. The corpus studies map the vulnerabilities disclosed after certification at scale \cite{janovsky2024seccerts}.

The second clock is the deployment's. E8 ended with administrators who never find the certified configuration; E9 adds the ones who reach it and then leave it, one reasonable decision at a time: a cipher suite enabled for a legacy client, a troubleshooting setting never switched back. Nothing announces the exit. There is no warning, no revocation, no light that turns off; the certificate reads the same on the day the configuration drifted as on the day it was earned.

Underneath both clocks, watching for drift is nobody's job: certification is an acquisition-time instrument, and everyone downstream treats it as an operations-time one.

The root cause is not negligence anywhere in the chain; it is the model. Common Criteria certifies a frozen pair, one version in one configuration, and both members start moving the moment the product ships; under continuous delivery there may be no stable version to freeze, which is where Section 7's open problems begin. Prevention, to the extent it exists, is in Section 6: patch-friendly architecture, guidance that states which updates preserve the evaluated configuration, and the self-attesting status check E8 asked for. None of this stops the clocks. It makes their movement observable. A certificate is most true on the day it is issued; every day after that belongs to E9.

\begin{table}[tp]
\caption{Nine failure modes of Common Criteria evaluation: where each surfaces, why it happens, and the Section 6 practice that prevents it.}\label{tab:modes}
\setlength{\tabcolsep}{3pt}\setlength{\aboverulesep}{0.3ex}\setlength{\belowrulesep}{0.45ex}\scriptsize
\newcolumntype{L}[1]{>{\raggedright\arraybackslash}p{#1}}
\begin{tabular}{@{}L{1.7cm}L{2.55cm}L{2.05cm}L{2.35cm}L{2.0cm}@{}}
\toprule
\hspace{0pt}Class & \hspace{0pt}The failure & \hspace{0pt}Surfaces at & \hspace{0pt}Root cause & \hspace{0pt}Prevention (\S6) \\
\midrule
\hspace{0pt}\textbf{E1} ST scoping & \hspace{0pt}Boundary too broad, too narrow, or hollowed by delegations & \hspace{0pt}ASE work units; churn before testing starts & \hspace{0pt}Scope set by market or consultant pressure before anyone checks that evidence can be produced & \hspace{0pt}Draw the boundary from evidence-production capability, before contract \\
\midrule
\hspace{0pt}\textbf{E2} PP conformance & \hspace{0pt}Product cannot deliver what the claimed PP demands & \hspace{0pt}Conformance-claim checks (ASE\_CCL) & \hspace{0pt}PP chosen by the vendor for market access before a gap analysis & \hspace{0pt}Requirement-by-requirement gap analysis against the exact PP version \\
\midrule
\hspace{0pt}\textbf{E3} Requirement coverage & \hspace{0pt}Claims conformant on paper, different on the wire & \hspace{0pt}Per-requirement test activities, mid-testing & \hspace{0pt}STs copied forward or written from spec sheets, not the shipping build & \hspace{0pt}Run the laboratory's enumeration against your own build first \\
\midrule
\hspace{0pt}\textbf{E4} Assurance evidence & \hspace{0pt}Documents miswritten, evidence untied, correct failures unshowable & \hspace{0pt}ADV/ALC work units; ATE\_IND evidence; ETR assembly & \hspace{0pt}Evidence treated as a byproduct of the work, not a product of it & \hspace{0pt}Write documents from the build; plan evidence capture before testing \\
\midrule
\hspace{0pt}\textbf{E5} Cryptography & \hspace{0pt}CAVP scope, entropy, key destruction & \hspace{0pt}CAVP verification; entropy assessment (SP 800-90B); FCS\_CKM & \hspace{0pt}Inheritance without ownership across silicon, library, and product & \hspace{0pt}Crypto inventory before claiming; start the entropy work first \\
\midrule
\hspace{0pt}\textbf{E6} Functional test coverage & \hspace{0pt}Testing touches less of the product than the certificate implies & \hspace{0pt}ATE\_IND / prescribed test activities & \hspace{0pt}Test effort follows testability, fixed by pre-evaluation design & \hspace{0pt}Drivable-as-client interfaces; inducible failure conditions \\
\midrule
\hspace{0pt}\textbf{E7} Vulnerability analysis & \hspace{0pt}Bounded survey read as unbounded adversarial testing & \hspace{0pt}AVA\_VAN & \hspace{0pt}Discovery is open-ended; the evaluation is bounded and discretionary & \hspace{0pt}Component inventory (SBOM); vendor-side CVE sweep; buy depth separately \\
\midrule
\hspace{0pt}\textbf{E8} Operational guidance & \hspace{0pt}Wrong steps, lost on reboot, or unread supplement & \hspace{0pt}AGD\_PRE / AGD\_OPE; bring-up & \hspace{0pt}Written away from the hardware; never tested like code & \hspace{0pt}Fresh-device walkthrough; explicit persistence; posture status command \\
\midrule
\hspace{0pt}\textbf{E9} Post-certification drift & \hspace{0pt}Product and deployment move; the certificate stands still & \hspace{0pt}After issuance; assurance continuity & \hspace{0pt}Point-in-time model applied to continuously changing systems & \hspace{0pt}Patch-friendly architecture; state config-preserving updates; self-attesting posture check \\
\bottomrule\end{tabular}\end{table}

\section{Cross-cutting analysis}
\secpagemark{Cross-cutting-analysis}

The taxonomy sorted failures by where they surface. This section reads it the other way, where four patterns repeat that no single class can show.

\subsection{The described product and the shipped one}

Strip the nine classes to their skeletons and the same shape appears in most of them: a description on one side, a behavior on the other, and the failure living in the distance between. E3 is the purest case, an ST that claims twelve cipher suites over a wire that shows nine. But the shape repeats at every scale. E1 and E2 are divergences between what the claim documents promise and what the organization can evidence or the build can deliver. E4 is a divergence between the work and its record. E8 is a divergence between the documented path and the device in the box; E9, between the certified pair and the fleet actually running. Even E7 fits, with the description living in the buyer's head: what ``vulnerability analysis'' is imagined to mean against what AVA\_VAN mandates.

An evaluation never examines a product alone; it examines a correspondence between a document set and a device, and the correspondence can break wherever its two sides are maintained by different people, different processes, or different companies (E5's chain crossed three). Anderson \& Fuloria and Murdoch et al. explain why divergence persists: the parties holding the information lack the incentive to close the gap \cite{anderson2009,murdoch2012}. The taxonomy locates where it lives: work unit by work unit, document by document. The two are diagnoses of the same disease at different resolutions, institutional and operational, and they compose. The incentive explains why nobody closes a given gap; the work unit is where someone could.

\subsection{Security is evaluated in the negative}

What an evaluator mostly does, hour by hour, is make bad things happen. The bench material behind this taxonomy is nearly all of that kind: certificate chains built to be broken, key types mismatched to cipher suites, oversized packets, wrong passwords, settings that must refuse to be set. The reason sits in what the requirements say. A security function is mostly a promise about what must not happen, so testing one means staging the forbidden thing and watching it be refused.

Three disciplines govern the staging, and each was met in an earlier class. The evaluator must be able to cause the failure: some failures can be staged only on a modified build (E6). The refusal must happen for the right reason: a device that rejects a broken chain with an unrelated error has not passed (E3). And the refusal must be shown, which is structurally harder than showing success: a completed handshake is its own evidence, while a correct rejection on an encrypted channel is a teardown without a visible cause (E4). Assembled, the pattern reads: negative claims are harder to stage, harder to judge, and harder to evidence than positive ones, and the certificate weighs both kinds the same. Section 6's requests for inducible failure conditions and verdict-showing interfaces are this pattern converted into design advice.

\subsection{The causes precede the evaluation}

Read down Table~\ref{tab:modes}'s root-cause column and ask when each cause happened. The boundary was set before the contract (E1). The PP was committed to before the gap analysis (E2). The ST was written from specification sheets before the build shipped (E3). The silicon and the library were licensed years earlier (E5). The interfaces that fix testability were designed before anyone thought about evaluation (E6). The guidance was written away from the hardware (E8). The evaluation is where failure surfaces, not where it originates, and the people who caused it were solving different problems and had never heard of the CEM.

A corollary is about ownership. Many of these early decisions are sourcing decisions, and each one scatters responsibility for the eventual claims across organizations that will never see the evaluation: E5's single ST vouching for silicon supplier, library maintainer, and product team at once; E7's survey blind to components nobody disclosed. Failures pool at the seams.

The exceptions are as instructive as the rule. E4's cause is a habit rather than a moment, and it spans both sides. E7 and E9 resist the pattern: their root causes are properties of the evaluation model, bounded analysis and point-in-time certification, not decisions any product team took. The distinction organizes the rest of the paper. Decisions can be made differently, which is Section 6's business; the model can only be changed, which is Section 7's.

\subsection{The certificate cannot say what it knows}

An evaluation ends knowing far more than its certificate can express. It knows which requirements got the thick end of the testing and which the thin (E6), how far the vulnerability survey actually reached (E7), the exact configuration in which the claims hold (E8), and the date after which nothing further is promised (E9). The certificate compresses all of it into one bit, certified, plus published documents that Section 2.3's buyers rarely open. Every buyer-side failure in the taxonomy runs through this compression: uneven assurance reads as uniform, a bounded survey reads as adversarial testing, an unreachable configuration reads as the product as shipped, and the exit from the evaluated state makes no sound.

None of this is deception. The information exists, in the ST, the Certification Report, and the ETR the buyer will never see; what is missing is bandwidth. The instrument that travels (the certificate, the listing, the word ``certified'') carries none of the texture, and the documents that carry the texture do not travel. The fixes inside the current model are channel fixes, and Section 6 collects them: a status command (E8, E9), a disclosed component inventory that makes the survey's reach checkable (E7). Whether the certificate could carry more than one bit, machine-readable claims or graded rather than binary verdicts, is taken up as an open problem in Section 7.

\section{Design-for-evaluability framework}
\secpagemark{Design-for-evaluability-framew}

Section 5.3 located the causes: decisions made before the laboratory was engaged, by people solving other problems. Prevention has to live in the same places, so this section is addressed to product teams, and every practice in it is unilateral: something one vendor can do alone, inside the current model, starting now.

The economic case is stronger than teams heading into a first evaluation tend to realize, because the exam is published. The Protection Profile names every requirement, and its test activities prescribe, step by step, what the laboratory will do to each one \cite{niapndcppsd}. Preparing is therefore bounded work: a known list, run once against one build. Failure has the opposite shape: billed re-test cycles, a firmware release, an amended ST that re-opens review, and calendar delay against the procurement the certificate was supposed to unlock \cite{janovsky2024seccerts,iccc2025fees}. Bounded, one-time cost on one side; unbounded, recurring cost on the other. The honest caveat: vendors without in-house security engineering often buy readiness to avoid costly stalls; the cheapest measure here costs nothing: knowing the playbook is public. Many first-time vendors do not.

The practices attach to three moments.

\subsection{When the claim is scoped}

\textbf{Draw the boundary from evidence, not ambition (E1).} For every component inside the proposed TOE boundary, name who will produce its design evidence and its test access. A component with no name attached is a stall already scheduled: move the line, or get the commitment in writing, before the contract fixes it.

\textbf{Run the gap analysis before committing to the PP (E2).} Requirement by requirement, against the exact PP version the market demands, against the shipping build rather than the roadmap. E2 called this the cheapest insurance an evaluation can buy: it is a spreadsheet and a week of one engineer's time, weighed against a product change discovered under contract.

\textbf{Inventory the cryptography before claiming it (E5).} Which library, which modes and key sizes, which CAVP certificate, earned on which operational environment. The claims inherit from this chain; the inventory is what makes them honest. And schedule the entropy assessment first, because it is the longest pole and the one least under the vendor's control.

\textbf{Inventory the components too (E7).} Third-party libraries and their versions, disclosed. The vulnerability survey searches what is named; the inventory gives it something to find.

\textbf{Decide what the certificate is for (E7).} When the product's real threat model exceeds basic attack potential, buy the deeper adversarial testing as its own engagement instead of reading it into a certificate that never promised it.

\subsection{What the design must provide}

Several classes trace to capabilities that either exist by the time evaluation starts or get improvised at the bench. The design provides them, or the evaluation pays for their absence.

\textbf{Make the device drivable as a client (E6):} a supported way to initiate protocol connections from its own interface, with test-controlled parameters, so that client-role requirements cost what server-role ones do.

\textbf{Make failures inducible (E6):} supported, documented ways to trigger the failure states that the requirements promise to handle, so that what gets tested is the shipping build and not a modified one.

\textbf{Make refusals speak (E4, Section 5.2):} when the device rejects something, have it say why, in a log line or error code that an evaluator or an administrator can capture. Success evidences itself; failure needs the product's help.

\textbf{Export the raw noise (E5):} a supported path to stream noise-source output at assessment volume off the device.

\textbf{Architect for the patch dilemma (E9):} structure updates so that security patches touch the evaluated functions as little as possible, which keeps assurance maintenance cheap enough to keep pace, and state in the guidance which classes of update preserve the evaluated configuration.

\textbf{Give the device a posture command (E8, E9).} Of everything in this section, this is the highest-leverage item. One supported command answers the question the back half of the taxonomy turns on: am I in the evaluated configuration right now? It repairs E8's unread supplement, because the device can say what the documents could not. It makes E9's silent drift visible to the people running the product. And it gives the buyer a way to check at operations time what the certificate could only promise at acquisition. Cryptographic modules have shipped approved-mode indicators for years \cite{fips1403}, and some vendors already ship automated tools that check whether a product is running in its certified configuration \cite{janovsky2024seccerts}; there is no technical reason a certified network device cannot answer for its own posture.

\subsection{Before the laboratory sees it}

\textbf{Run the laboratory's tests first (E3).} The per-requirement test activities are published, and the enumeration they prescribe runs on commodity tooling from any bench. What it catches is exactly E3: stale claims, spec-sheet claims, features the hardening guide disables, at desk cost instead of contract cost.

\textbf{Sweep the inventory for CVEs (E7),} before the laboratory does, so that whatever the survey will find is already known, and either patched or answered.

\textbf{Walk a fresh device through the guidance (E8).} Someone who did not build the product and did not write the documents takes a fresh unit from the box to the evaluated configuration using the draft guidance alone, with the persistence step explicit at every point it matters. Every stumble is a bug. This is the only testing that documentation ever gets, and the evaluator should be the second person to make that walk, not the first.

\textbf{Write the documents from the build, and plan the evidence (E4).} The functional specification and the configuration-management evidence describe the build being shipped to the laboratory. Before testing starts, decide how each outcome will be observed and recorded, especially each intended failure, and keep the artifacts in a tree that mirrors the report that will carry them.

None of this is exotic. It is the discipline that software teams already apply to code (tests before release, dependency inventories, documentation that gets executed) applied to the artifacts an evaluation consumes. It earns a section in a research paper because of Section 5.3's finding: these artifacts are routinely produced by people who never see the bench, under assumptions the bench does not honor.

What the framework cannot do is worth stating just as plainly. The practices above prevent the seven decision-caused classes; they only soften the two structural ones. An honest component inventory makes E7's survey better without making it unbounded, and a posture command makes E9's drift visible without making the certificate current. What remains after prevention is the model itself, and the model is Section 7's subject.

\section{Open problems}
\secpagemark{Open-problems}

Section 6's practices stop where the model itself is the cause; the problems below live past that boundary. Like E9, this section reasons from the public record and the model's structure; the bench appears only where E5 left off. Each problem is stated with what makes it hard and what a solution would need to provide.

\subsection{Certifying software that moves}

E9 named the frozen pair and the two clocks that run against it. Continuous delivery breaks the first member of the pair: under weekly or daily release trains, a months-long evaluation certifies a build that is archaeology by the time the certificate issues. Cloud deployment breaks the second: a SaaS product has no customer-side evaluated configuration to reach, because the vendor operates the only deployment there is, and operates it as one continuously changing fleet. Assurance continuity was built for neither \cite{ccraac2012}, and the newer mechanisms at the margins still anchor to a named version \cite{eucc2024}. The open question is what the unit of certification becomes when the product is a stream rather than an artifact. Certify the pipeline and process that produce the builds, sampling builds for spot evaluation? Certify a smaller, stabler core and accept a moving periphery? Each candidate trades away part of what a certificate now means, and neither has a scheme-workable version yet.

\subsection{Requirements for components that cannot be enumerated}

The PP-based model's precision rests on enumerability. A claim is a list (these suites, these versions, these key sizes) and the test activity walks the list one entry at a time. An ML component breaks the assumption at every step. Its behavior is a learned function, not a list; its version changes when the weights do, which collides with 7.1; and its characteristic failures (adversarial inputs, distribution shift) are discovered rather than specified, E7's bounded-analysis problem with the bound removed. It is not clear what an SFR for a classifier should assert, or what a prescribed test activity for one would look like. Early work approaches the maintenance side \cite{lopezmunoz2025}; the requirements side, what to claim and how to check it, is open.

\subsection{Sizing evidence demands to the hardware}

E5 met the mild version of this problem: an entropy assessment that needs more consecutive raw output than the device can store \cite{sp80090b}, satisfied through improvised collection plumbing. Virtualized platforms pose the hard version. A virtual machine's randomness arrives through layers the tenant cannot inspect, from hardware the tenant does not control, so the assessment's first question, where does the randomness physically come from, may have no answer available to the party being evaluated. The open problem is methodology whose evidence demands fit the platforms being assessed: entropy validation for constrained devices, and any defensible assessment model for virtualized noise sources.

\subsection{A certificate that can say more}

Section 5.4 diagnosed the instrument: one bit, plus documents that do not travel. The open problem is whether the instrument can carry more without breaking what it already does. The taxonomy supplies the candidate content: where testing ran thick and where it ran thin (E6), how far the vulnerability survey reached (E7), and above all the evaluated configuration itself, published today as prose in a supplement --- it could be a machine-readable, checkable artifact. That last candidate is the standardized counterpart of Section 6's per-device posture command, and it would also address guidance that drifts across vendors and re-certifications (E8). The demand side is already proven: researchers mine the certificate corpus at scale and fight its unstructured form to do it \cite{janovsky2024seccerts}. The supply side is the open half, and the harder part of it is not format design but behavior: whether procurement, handed more bits, would read them.

\subsection{One arrangement, diverging methods}

This paper leaned on the United States scheme's exact-conformance model; much of the world still evaluates against EALs, and Europe now operates its own framework \cite{eucc2024}. The divergence is methodological, not cosmetic. Exact conformance buys comparability within a product category and gives up graded depth; EALs offer graded depth and give up cross-product comparability; and the recognition arrangement holding the system together already caps what it recognizes across borders \cite{ccra2014}. Whether the tracks converge, coexist under narrowing mutual recognition, or fragment into regional regimes is open, and the answer prices what ``internationally recognized'' is worth.

\section{Related work}
\secpagemark{Related-work}

Work bearing on Common Criteria evaluation falls into the three strands Section 1 introduced, plus several supporting literatures.

\textbf{Economics of certification.} The institutional critique begins with Anderson and Fuloria, who analyze certification as an asymmetric-information market, argue that misaligned incentives can reduce it to security theatre, and sketch what they explicitly call a rough taxonomy of how evaluation and certification fail \cite{anderson2009}. Murdoch, Bond and Anderson extend the critique: the vendor chooses and pays the laboratory, the conflict of interest is structural, and the PIN-entry-device case of Section 1, complete with a certification report withheld from relying parties, is their standing example \cite{murdoch2012}. This strand is the taxonomy's closest neighbor, and the relationship is inheritance rather than overlap. Their failure modes are institutional (who pays, who knows, who answers for it) and deliberately cross-domain; ours are operational and CC-specific, located in the work units and evidence of a single evaluation. Section 5.1 argued that the two compose: the incentives explain why a divergence persists, and the work units are where it lives.

\textbf{Certificate-corpus measurement.} The sec-certs line mines the public record at scale, mapping disclosed vulnerabilities onto certified products and testing whether assurance levels predict them \cite{janovsky2024seccerts}. Corpus measurement necessarily treats each evaluation as a unit; this paper is qualitative and opens the unit up. Where their data touches the taxonomy (vulnerabilities surfacing after certification, E7 and E9) we cite it as the quantitative record of failures whose mechanics we describe.

\textbf{Adoption and requirements surveys.} A third strand surveys the standard itself: how organizations define security requirements with CC and what obstructs its adoption \cite{sun2022}, and how the national evaluation schemes compare \cite{fatima2021}. These works are pro-adoption in orientation, asking how to use the framework well; ours asks why evaluations under it fail.

\textbf{Practitioner evidence.} Several of the taxonomy's authorship claims rest on public practitioner sources rather than academic literature. An ICCC analysis by a laboratory technical director documents that third-party consultants are frequently utilized and author Security Targets and guidance documents \cite{arnold2006,ewa2006}, and a dedicated CC-documentation industry has operated for over two decades \cite{rosenberry}. E1 and E8 lean on this record where they attribute authorship.

\textbf{Certification under continuous change.} The nearest current work to problems 7.1 and 7.2 comes at them from the maintenance side: machinery that keeps certification evidence current for learning-enabled systems under continuous change, rather than a failure taxonomy \cite{lopezmunoz2025}.

\textbf{Historical lineage.} Retrospectives on TCSEC and the first-generation evaluation criteria record complaints with a familiar ring: evaluations too slow for product cycles, costs concentrated on vendors, certified configurations misaligned with what users actually run \cite{lipner2015,nistir4774}. The same themes recurring across forty years and three assurance regimes suggest the failure modes of Section 4 are properties of the evaluation-certification form rather than artifacts of the current standard.

\textbf{Assurance cases and design for assurance.} The security-assurance-case literature develops structured argumentation for security claims, largely in the safety-critical tradition \cite{mohamad2021}. Section 6 shares the premise, that assurance is cheaper designed in than bolted on, and narrows it: the framework's practices attach to specific CC work units and failure classes rather than to general argument structure.

Across these strands we find no prior systematization of Common Criteria evaluation failure from the evaluator's operational vantage, at the level of work units, test activities, and evidence. The economics literature names the incentives, the corpus literature measures the outcomes, and the surveys map adoption; the inside of the evaluation, where failure surfaces, has not previously been the object of study. That is the gap this paper fills.

\section{Conclusion}
\secpagemark{Conclusion}

Prior work established that Common Criteria certification has an incentive problem and a measured outcome problem. This paper opened the box between the two: the evaluation itself, systematized from the chair of the person who runs it. The result is a taxonomy of nine failure classes spanning the lifecycle, from a boundary drawn before any contract is signed to a certificate aging after issuance, each anchored to the work units where an evaluator meets it, each traced to a root cause, and each paired with the practice that prevents it.

The constructive finding is the one we most want practitioners to keep: most of this is preventable, cheaply and unilaterally. Seven of the nine classes trace to decisions a product team controls, the laboratory's playbook is public, and every practice in Section 6 can be adopted by one vendor alone. Design for evaluability asks for no new discipline, only for existing engineering discipline to reach the artifacts an evaluation consumes.

What prevention cannot reach is the model. Section 7's problems mark where the strain shows, and product practice is moving toward the strain, not away.

We wrote this systematization so that the field's two mature literatures, the incentives and the outcomes, would have their missing middle: the mechanics. And for a simpler reason, visible from the bench: so that the next person unboxing a router in a laboratory finds a device that was built to be evaluated, and guidance that was written to be followed.

%% file: fig1.tex
\resizebox{\textwidth}{!}{%
\begin{tikzpicture}[
  cls/.style={draw, align=center, inner sep=2.5pt, minimum height=8mm, font=\scriptsize},
  mdl/.style={cls, fill=black!12},
  ph/.style={font=\itshape\scriptsize, align=center},
  flow/.style={-{Stealth[length=2.2mm]}, semithick}]
\node[cls, minimum width=2.1cm]  (e1) at (1.05, 0.46) {E1 ST scoping};
\node[cls, minimum width=2.1cm]  (e2) at (1.05,-0.46) {E2 PP conformance};
\node[cls, minimum width=2.05cm] (e3) at (3.375, 0.46) {E3 Requirement\\coverage};
\node[cls, minimum width=2.05cm] (e4) at (3.375,-0.46) {E4 Assurance\\evidence};
\node[cls, minimum width=2.05cm] (e5) at (5.505, 0.46) {E5 Cryptography};
\node[cls, minimum width=2.05cm] (e6) at (5.505,-0.46) {E6 Test coverage};
\node[mdl, minimum width=1.95cm] (e7) at (7.755,0) {E7 Vulnerability\\analysis};
\node[cls, minimum width=1.9cm]  (e8) at (9.93,0) {E8 Operational\\guidance};
\node[mdl, minimum width=2.1cm]  (e9) at (12.38,0) {E9 Post-\\certification drift};
\draw[flow] (2.12,0) -- (2.33,0);
\draw[flow] (6.55,0) -- (6.76,0);
\draw[flow] (8.75,0) -- (8.96,0);
\draw[flow] (10.9,0) -- (11.31,0);
\draw[line width=1.1pt] (11.11,-0.95) -- (11.11,1.9);
\node[font=\scriptsize\bfseries, align=center] at (11.11,2.2) {certificate\\issued};
\node[ph] at (1.05,1.25)  {Scoping \&\\conformance};
\node[ph] at (4.44,1.25)  {Evidence, cryptography\\\& functional testing};
\node[ph] at (7.755,1.25) {Vulnerability\\analysis};
\node[ph] at (9.93,1.25)  {Guidance:\\the last mile};
\node[ph] at (12.38,1.25) {Post-\\certification};
\draw[{Stealth[length=1.8mm]}-{Stealth[length=1.8mm]}] (0,1.9) -- node[fill=white, font=\scriptsize\itshape, inner sep=1.5pt]{the evaluation} (10.88,1.9);
\draw[{Stealth[length=1.8mm]}-{Stealth[length=1.8mm]}] (11.33,1.9) -- node[fill=white, font=\scriptsize\itshape, inner sep=1.5pt]{the field} (13.43,1.9);
\draw[flow] (0,-1.15) -- (13.43,-1.15);
\draw (2.7,-1.72) rectangle +(0.3,0.22);
\node[anchor=west, font=\scriptsize] at (3.06,-1.61) {decision-caused, preventable (Section 6)};
\draw[fill=black!12] (8.35,-1.72) rectangle +(0.3,0.22);
\node[anchor=west, font=\scriptsize] at (8.71,-1.61) {model-caused, open problems (Section 7)};
\end{tikzpicture}}

%% file: appendixA.tex
\appendix

\section{Evidence Anchors by Failure Class}

This appendix consolidates the public-record anchors behind Section 4's taxonomy so that each class's evidentiary basis can be checked independently: one row per anchor, with the work units it attaches to. The work units are the CEM's and are common to EAL-graded and exact-conformance evaluations. As Section 3 discloses, the classes differ in basis: for E1, E2, E5, E7, and E9 the public record carries the class; for E3, E4, E6, and E8 the anchors corroborate mechanics that generalized bench practice supplies. No row is a claim about frequency.

\begin{center}
\footnotesize
\setlength{\tabcolsep}{4pt}
\renewcommand{\arraystretch}{1.15}
\begin{tabular}{@{}l>{\raggedright\arraybackslash}p{3.05cm}>{\raggedright\arraybackslash}p{2.15cm}>{\raggedright\arraybackslash}p{5.15cm}@{}}
\toprule
\textbf{Class} & \textbf{Anchor} & \textbf{Work-unit locus} & \textbf{What it evidences} \\
\midrule
E1 & ST 383-7-138, v2.1, revision history \cite{st3837138} & ASE & Boundary redrawn mid-evaluation: ``Expanded the TOE boundary to address certifier ORs'' \\
E1 & JBoss EAP 4.3.0 ST, BSI-DSZ-CC-0531 \cite{bsi0531} & ASE & Revisions log the boundary moving in both directions: requirements removed, TOE components added \\
E2 & TD0592 \cite{td0592} & ASE\_CCL & PP-mandatory local audit storage absent from a shipping product; resolved by scheme decision \\
E2 & TD0009 \cite{td0009} & ASE\_CCL & Claimed cipher-suite selections in conflict with what implementations honor (WLAN AS PP) \\
E3 & FortiMail ST v1.13; ST VID 11253 \cite{st3834314,stvid11253} & ATE, prescribed activities \cite{niapndcppsd} & Amended STs on the record: testing halts on the affected requirements, the claim is amended, review re-opens \\
E4 & Practitioner FAQ \cite{jtsecfaq} & CEM document checks & Laboratory--vendor iteration until each document carries the CEM's checks, documented as the working norm \\
E4 & ICCC practitioner reports \cite{arnold2006,ewa2006} & ADV, ALC inputs & Documents written for the wrong job, reported from the evaluation floor \\
E5 & NIAP Policy Letter \#5 \cite{niappolicy5} & FCS; CAVP gate & Every claimed algorithm with a NIST validation program must hold a CAVP certificate; scope mismatches are hard stops \\
E5 & SP 800-90B \cite{sp80090b} & Entropy assessment & Raw-data demands that exceed what shipping devices store or expose \\
\bottomrule
\end{tabular}
\end{center}

\begin{center}
\footnotesize
\setlength{\tabcolsep}{4pt}
\renewcommand{\arraystretch}{1.15}
\begin{tabular}{@{}l>{\raggedright\arraybackslash}p{3.05cm}>{\raggedright\arraybackslash}p{2.15cm}>{\raggedright\arraybackslash}p{5.15cm}@{}}
\toprule
\textbf{Class} & \textbf{Anchor} & \textbf{Work-unit locus} & \textbf{What it evidences} \\
\midrule
E5 & TD0057 $\rightarrow$ TD0130 \cite{td0057,td0130} & FCS\_CKM key destruction & Public iteration arc on one requirement: zeroization true at the API, false on wear-leveled media \\
E6 & NDcPP Supporting Document \cite{niapndcppsd} & ATE\_IND activities & The prescribed per-requirement test activities whose nominal completeness E6 measures against \\
E6 & TD0190 \cite{td0190} & ATE\_IND & Failure-state requirements that cannot be induced without vendor test builds \\
E7 & CEM \cite{cem2022} & AVA\_VAN & The actual mandate: public-source survey plus penetration testing at basic attack potential---narrower than buyers read \\
E7 & ROCA \cite{nemec2017} & AVA\_VAN bound & Factorable keys in certified smartcards, found by researchers outside any evaluation \\
E7 & PIN-device attacks \cite{murdoch2012} & AVA\_VAN bound & Inexpensive attacks on certified payment devices, demonstrated post-certification \\
E7, E9 & Certificate corpus \cite{janovsky2024seccerts} & Post-certification record & Disclosed vulnerabilities mapped onto certified products at scale \\
E8 & CC documentation services \cite{rosenberry,arnold2006} & AGD & Guidance authored away from the hardware by writers and consultants---industry practice on the record since 2001 \\
E9 & CCRA assurance continuity \cite{ccraac2012} & Maintenance & Version-anchored maintenance: the mechanism E9 measures against exploit-speed patching \\
\bottomrule
\end{tabular}
\end{center}

%% file: references.bib
@misc{cem2022,
  author = {{CCRA}},
  title = {Common Methodology for Information Technology Security Evaluation ({CEM}), {CC}:2022 Revision 1 ({ISO/IEC} 18045)},
  year = {2022}
}

@misc{niapndcpp,
  author = {{NIAP}},
  title = {collaborative {P}rotection {P}rofile for {N}etwork {D}evices ({NDcPP}), v2.2e},
  year = {2020}
}

@misc{niapndcppsd,
  author = {{Network iTC}},
  title = {Supporting Document: Evaluation Activities for Network Device {cPP}},
  year = {2020},
  note = {Version 2.2}
}

@misc{ccra2014,
  author = {{CCRA}},
  title = {Arrangement on the Recognition of Common Criteria Certificates in the Field of Information Technology Security},
  year = {2014}
}

@misc{ccraac2012,
  author = {{CCRA}},
  title = {Assurance Continuity: {CCRA} Requirements, Version 2.1},
  year = {2012}
}

@misc{eucc2024,
  author = {{European Commission}},
  title = {Commission Implementing Regulation ({EU}) 2024/482 on the {E}uropean {C}ommon {C}riteria-based cybersecurity certification scheme ({EUCC})},
  year = {2024},
  note = {{OJ} L, 7 February 2024}
}

@misc{cnssp11,
  author = {{CNSS}},
  title = {National Policy Governing the Acquisition of Information Assurance ({IA}) and {IA}-Enabled Information Technology Products},
  year = {2013},
  note = {{CNSS} Policy No.~11}
}

@misc{niappolicy5,
  author = {{NIAP}},
  title = {Policy Letter \#5, with Addenda 1--2},
  note = {\url{https://www.niap-ccevs.org/Policy/}}
}

@techreport{sp80090b,
  author = {Turan, Meltem S{\"o}nmez and others},
  title = {Recommendation for the Entropy Sources Used for Random Bit Generation},
  institution = {NIST},
  number = {SP 800-90B},
  year = {2018}
}

@misc{fips1403,
  author = {{NIST}},
  title = {Security Requirements for Cryptographic Modules ({FIPS} 140-3)},
  year = {2019}
}

@article{murdoch2012,
  author = {Murdoch, Steven J. and Bond, Mike and Anderson, Ross},
  title = {How Certification Systems Fail: Lessons from the {W}are Report},
  journal = {IEEE Secur. Priv.},
  volume = {10}, number = {6}, pages = {40--44}, year = {2012},
}

@inproceedings{anderson2009,
  author = {Anderson, Ross and Fuloria, Shailendra},
  title = {Certification and Evaluation: A Security Economics Perspective},
  booktitle = {IEEE ETFA},
  year = {2009},
}

@article{janovsky2024seccerts,
  author = {Janovsk{\'y}, Adam and others},
  title = {sec-certs: Examining the security certification practice for better vulnerability mitigation},
  journal = {Comput. Secur.},
  volume = {143}, year = {2024},
}

@article{sun2022,
  author = {Sun, Nan and others},
  title = {Defining Security Requirements With the {C}ommon {C}riteria: Applications, Adoptions, and Challenges},
  journal = {IEEE Access},
  volume = {10}, pages = {44756--44777}, year = {2022}
}

@article{fatima2021,
  author = {Fatima, Anum and others},
  title = {A survey on common criteria ({CC}) evaluating schemes for security assessment of {IT} products},
  journal = {PeerJ Comput. Sci.},
  volume = {7}, pages = {e701}, year = {2021}
}

@misc{lopezmunoz2025,
  author = {Lopez-Flores, S. and Mu{\~n}oz, A.},
  title = {Evidence-Centric Certification Maintenance for Learning-Enabled Systems Under Continuous Change},
  year = {2025},
  note = {{Research Square} preprint rs-8967881/v1}
}

@article{lipner2015,
  author = {Lipner, Steven B.},
  title = {The Birth and Death of the {O}range {B}ook},
  journal = {IEEE Ann. Hist. Comput.},
  volume = {37}, number = {2}, pages = {19--31}, year = {2015}
}

@techreport{nistir4774,
  author = {Dinkel, Charles},
  title = {A Review of {U.S.} and {E}uropean Security Evaluation Criteria},
  institution = {NIST},
  number = {NISTIR 4774},
  year = {1992}
}

@article{mohamad2021,
  author = {Mohamad, Mazen and others},
  title = {Security assurance cases---state of the art of an emerging approach},
  journal = {Empir. Softw. Eng.},
  volume = {26}, year = {2021},
}

@inproceedings{nemec2017,
  author = {Nemec, Matus and others},
  title = {The Return of {C}oppersmith's Attack: Practical Factorization of Widely Used {RSA} Moduli},
  booktitle = {ACM CCS},
  year = {2017},
}

@inproceedings{moghimi2020,
  author = {Moghimi, Daniel and Sunar, Berk and Eisenbarth, Thomas and Heninger, Nadia},
  title = {{TPM-FAIL}: {TPM} meets Timing and Lattice Attacks},
  booktitle = {USENIX Security},
  year = {2020}
}

@article{jancar2020,
  author = {Jan{\v c}{\'a}r, J{\'a}n and others},
  title = {Minerva: The curse of {ECDSA} nonces},
  journal = {IACR TCHES},
  volume = {2020}, number = {4}, pages = {281--308}, year = {2020}
}

@inproceedings{arnold2006,
  author = {Arnold, James},
  title = {Impacts of Third Party Consultants on {C}ommon {C}riteria Assurances},
  booktitle = {7th International Common Criteria Conference (ICCC)},
  year = {2006},
  note = {{Lanzarote}, Spain}
}

@inproceedings{ewa2006,
  author = {Connor, E. and Gauvreau, M. and Moore, S. E.},
  title = {Developer Documentation: A Who To Guide},
  booktitle = {7th ICCC},
  year = {2006},
  note = {{EWA-Canada}; {Lanzarote}, Spain}
}

@misc{rosenberry,
  author = {{Rosenberry Associates, Inc.}},
  title = {{FIPS} and {C}ommon {C}riteria documentation services},
  note = {\url{https://www.raisec.com}, developing CC documentation since 2001; accessed 2026}
}

@inproceedings{iccc2025fees,
  author = {{Binti Jeffridin}, N.},
  title = {Evaluation cost data for the {MyCC} scheme (session {A22c})},
  booktitle = {ICCC},
  year = {2025},
}

@misc{jtsecfaq,
  author = {{jtsec Beyond IT Security}},
  title = {{C}ommon {C}riteria {FAQ}},
  note = {\url{https://www.jtsec.es} (2026)}
}

@misc{td0009,
  author = {{NIAP}},
  title = {{TD0009}: {WLAN} {AS} {PP} cipher-suite conflict resolution},
  year = {2015},
}

@misc{td0057,
  author = {{NIAP}},
  title = {{TD0057}: Key destruction on wear-leveled flash memory},
  year = {2015},
}

@misc{td0130,
  author = {{NIAP}},
  title = {{TD0130}: Destruction of cryptographic keys ({NDcPP})},
  year = {2016}
}

@misc{td0190,
  author = {{NIAP}},
  title = {{TD0190}: Failure states requiring {TSF} shutdown},
  year = {2017},
}

@misc{td0592,
  author = {{NIAP}},
  title = {{TD0592}: Local storage of audit records},
  year = {2021}
}

@misc{st3837138,
  author = {{Canadian Common Criteria Scheme}},
  title = {Security Target 383-7-138, v2.1},
  year = {2015},
  note = {Revision history: ``Expanded the TOE boundary to address certifier ORs''}
}

@misc{bsi0531,
  author = {{Red Hat}},
  title = {Security Target for {JBoss} {E}nterprise {A}pplication {P}latform 4.3.0 ({BSI-DSZ-CC-0531})},
  year = {2009}
}

@misc{st3834314,
  author = {{Fortinet}},
  title = {{FortiMail} Security Target, v1.13 (383-4-314)},
  year = {2015}
}

@misc{stvid11253,
  author = {{NIAP}},
  title = {Security Target, {VID} 11253},
}
